\documentclass[12pt,preprint]{aastex}

\usepackage{emulateapj5}
\usepackage{graphics}
\usepackage{psfig}

\def\93j{{SN~1993J}}
\def\R{{\sl ROSAT}}
\def\X{{\sl XMM-Newton}}

\def\C{{\sl Chandra}}
\def\B{{\sl BeppoSax}}

\begin{document}

\title{X-ray Emission from the Type Ic Supernova 1994I 
Observed with {\sl Chandra}}

\author{Stefan Immler}
\affil{Astronomy Department, University of Massachusetts, Amherst, MA 01003}
\author{Andrew S. Wilson \& Yuichi Terashima\altaffilmark{1}}
\affil{Astronomy Department, University of Maryland, College Park, MD 20742} 
\altaffiltext{1}{Institute of Space and Astronautical Science, 3-1-1 Yoshinodai, 
Sagamihara, Kanagawa 229-8510, Japan}

\shorttitle{X-ray emission from the Type Ic SN~1994I}
\shortauthors{Immler, Wilson \& Terashima}

\begin{abstract}
We present two high-resolution \C\ X-ray observations of supernova (SN) 1994I 
which show, for the first time, that the interaction of the blast wave from a Type Ic 
SN with its surrounding circumstellar material (CSM) can give rise to soft X-ray 
emission. Given a 0.3--2~keV band 
X-ray luminosity of $L_{\rm x}\sim1\times10^{37}~{\rm ergs~s}^{-1}$ between six and 
seven years after the outburst of SN~1994I, and assuming the X-ray emission arises 
from the shock-heated CSM, we derive a pre-SN mass-loss rate of
$\dot{M}\sim1\times10^{-5}~M_{\odot}~{\rm yr}^{-1}~(v_{\rm w}/10~{\rm km~s}^{-1})$. 
Combining the results with earlier \R\ observations, we construct the X-ray lightcurve 
of SN~1994I. A best-fit X-ray rate of decline of $L_{\rm x} \propto t^{-s}$ with
index $s\sim1$ and a CSM density profile of $\rho_{\rm csm} \propto r^{-1.9\pm0.1}$ 
are inferred, consistent with what is expected for a constant mass-loss rate and 
constant wind velocity profile for the SN progenitor ($\rho_{\rm csm} \propto r^{-2}$).
\end{abstract}


\section{Introduction}
\label{introduction}

The classification scheme for SNe is based on the presence or absence of 
hydrogen lines in their optical spectra (Type II and Type I, respectively). While Type 
II/Ib SNe occur during the core collapse of a massive star, Type Ia SNe are believed to 
originate from the deflagration or detonation of an accreting white dwarf star in a binary 
system. The massive ($>10 M_{\odot}$) progenitor stars for the controversial Type Ic SN 
subclass are thought to have lost their outer hydrogen and helium layers either through 
stellar winds or by mass transfer to a companion, leaving behind a stripped carbon and 
oxygen (C+O) star. Such a `naked' C+O star will later explode when its iron core 
collapses (e.g. Nomoto et al. 1994). 
The interaction of the outgoing SN shock wave with the ambient CSM, 
deposited either by a pre-SN stellar wind or non-conservative mass transfer to a 
companion, produces hot gas with characteristic temperatures in the range 
$T \sim 10^7$--$10^9$~K (Chevalier \& Fransson 1994). 
Gas heated to such high temperatures produces radiation predominantly in the X-ray 
range. X-ray emission from this interaction is expected for all Type Ib/c and II SNe with 
substantial CSM established by the massive progenitors. Over the last 20 years, searches 
for X-ray emitting SNe have been successful for only a relatively small number of Type 
II's in the near aftermath (days to months) of the explosion (see Immler \& Lewin 2002 
for a review article)\footnote{A complete list of 
X-ray SNe is available at http://xray.astro.umass.edu/sne.html}.

By contrast, no Type Ia or Ib/c SN has ever been firmly detected in X-rays,
apart from a $3.5\sigma$ excess in an \X\ EPIC-PN image $4''$ offset from the optical 
position of the Ic SN~2002ap (Rodriguez-Pascual et al. 2002). Hard (2--10~keV band) X-ray 
emission was recorded with \B\ from the position of the unusual SN 1998bw, which might 
be associated with a $\gamma$-ray burst event (GRB 980425; Galama et al. 1998, 
Pian et al. 1998). Due to the large error box of the \B\ Wide Field Camera observation
(between $3'$ and $8'$, $99\%$ confidence limit) and the rather large probability of 
$\sim60\%$ that the source is a random chance coincidence, the association of the 
X-ray source with SN 1998bw and GRB~980425 is still tentative (Galama et al. 1998).

Evidence for soft (0.1--2.4~keV band) X-ray emission has been reported from
the Type Ic SN~1994I, based on \R\ HRI observations 82 days after the 
outburst (Immler, Pietsch \& Aschenbach 1998). 
However, the \R\ observations, with a spatial resolution of $\sim5''$ (FWHM 
on-axis), were not conclusive since SN~1994I is located close to the X-ray bright nucleus 
of the host galaxy M51 (distance $\sim18''$) and is embedded in a high level of extended 
X-ray emission from hot gas and unresolved point-like X-ray sources in the bulge of M51. 

\section{X-Ray Observations and Analysis}
\label{obs}
We used the \C\ X-ray observatory to search for X-ray emission from SN 
1994I in X-ray images taken between six (June 6, 2000) and seven years (June 23, 2001)  
after the explosion (March 31, 1994; Chandler, Phillips \& Rupen 1994). 
Our 14.9~ks and 26.8~ks \C\ observations 
were carried out with the S3 chip of the Advanced CCD Imaging Spectrometer (ACIS-S) 
at the focal plane of the telescope and the nucleus of the host galaxy M51 placed 
at the aim-point (Terashima \& Wilson 2001). 
The superb spatial resolution ($0\farcs5$ FWHM on-axis), together with the 
high sensitivity of the instrument and the long combined exposure (41.7~ks), allows us to 
separate point-like sources from the diffuse emission in the bulge of M51 and to carry 
out a sensitive search for X-ray emission from SN~1994I. The co-added and adaptively 
smoothed \C\ image of the central region of M51 is presented in Fig.~1.

An X-ray source is detected at the position of SN~1994I in the merged \C\ data with 
a significance of $\sim6\sigma$ in the 0.3--2~keV band using the `wave detect' algorithm 
implemented in the {\sc ciao} data analysis package. The position of the X-ray source 
(${\rm R.A., Dec.} (2000) = 13^{\rm h} 29^{\rm m} 54\fs17, +47\degr 11' 30\farcs2$) 
is fully consistent with the radio position of the SN ($\sim0\farcs5$ offset; 
Rupen et al. 1994). Astrometry of the M51 nucleus shows a similarly small offset 
between the (2--8~keV band) X-ray and radio positions ($\sim0\farcs3$). Exposure corrected 
source counts were extracted within a radius of 3 image pixels (90\% encircled energy 
radius at 1~keV) and corrected for the background taken in an annulus with inner and 
outer radii of 3.5 and 9.5 pixels, respectively (1 pixel corresponds to $0\farcs49$).

A spectral model has to be assumed to convert source counts into energy fluxes. 
We adopted an effective (0.3--2~keV band) cooling function of  
$\Lambda = 3 \times 10^{-23}~{\rm ergs~cm}^3~{\rm s}^{-1}$ 
for an optically thin Raymond-Smith thermal plasma with a temperature of 
$10^7$~K (Raymond, Cox \& Smith 1976). Although the temperature is unknown, this
value is consistent with theoretical expectations (Chevalier \& Fransson 1994), 
previous X-ray observations of other SNe and the detection 
of SN~1994I in the soft (0.3--2~keV) X-ray band where the peak of a $10^7$~K 
spectrum is located. The equivalent count rate to (unabsorbed) flux conversion factors 
are then $4\times10^{-11}~({\rm ergs~cm}^{-2}~{\rm s}^{-1})/({\rm counts~s}^{-1})$ 
and $3\times10^{-12}~({\rm ergs~cm}^{-2}~{\rm s}^{-1})/({\rm counts~s}^{-1})$ 
for the \R\ HRI and \C\ ACIS-S3, respectively, and a Galactic foreground column 
density of $N_{\rm H} = 1.3\times10^{20}~{\rm cm}^{-2}$ (Dickey \& Lockman 1990). 
The uncertainty of the conversion factors for optically thin thermal spectra with 
temperatures in the range $10^7$--$10^9$~K is $\sim15\%$. Assuming a 0.86~keV thermal 
bremsstrahlung spectrum (corresponding to a temperature of $10^7$~K) instead of a 
Raymond-Smith thermal plasma increases the conversion factor by $\sim7\%$. 
The \C\ results for the two individual observations of SN~1994I, together with 
previous \R\ HRI results (Immler, Pietsch \& Aschenbach 1998), are summarized in Table~1.

\section{Discussion}
\label{discussion}

We used the \C\ and \R\ data to construct a combined X-ray lightcurve 
of SN~1994I, which is presented in Fig.~2. A best-fit single power-law rate 
of decline of $f_{\rm x} \propto t^{-s}$ with index $s=0.9^{+0.1}_{-0.2}$ 
(solid line, Fig.~\ref{f2}) is inferred for the measurements on days 82, 2,271 and 
2,639. Assuming an initial exponential rise of the X-ray luminosity after the 
outburst (at time $t_0$) of SN~1994I and a subsequent power-law decline with 
index $s$, we also parameterized the X-ray evolution as 
$f_{\rm x} \propto (t-t_0)^{-s} \times e^{-\tau}$ with $\tau \propto (t-t_0)^{-\beta}$. 
This model has been successfully used to describe the time dependence of the 
radio emission of SNe (Weiler et al. 1996).
The external absorption of the emission is represented by the $e^{-\tau}$
term (`optical depth') and the time-dependence of the optical depth is parameterized 
by the exponent $\beta$. In the X-ray regime, the rise could represent either 
decreasing absorption by material along the line of sight to the hot gas or 
simply non-production of X-rays in the \R\ band at early times.
If we adopt this strictly heuristic description, we find $s\sim1$ and $\beta\sim2$
(dashed line, Fig.~\ref{f2}). In case the \R\ detection on day 82 corresponds to a 
pre-maximum measurement, a steeper index of $s\sim1.5$ is inferred 
(dotted line, Fig.~\ref{f2}). However, given the large error range of the flux
measurements, $s$ and $\beta$ are not very well constrained.

While a $t^{-1}$ rate of decline has been observed for the Type IIb SN~1999em with
\C\ (Pooley et al. 2002), faster rates of decline have been reported for the Ic SN 1998bw 
($s=1.4$), inferred from the tentative hard (2--10~keV) band \B\ data (Pian et al. 1999), 
the Ic radio SN~1990B ($s=1.3$) and the Ib radio SNe 1983N ($s=1.6$) and 1984L
($s=1.5$; Van Dyk et al. 1993). By contrast, the long-term X-ray lightcurve
of the IIb SN~1993J is best described by a slow rate of decline with $s=0.27$
(Immler, Aschenbach \& Wang 2001).

The X-ray emitting material could be either shocked SN ejecta (reverse shock) or 
shocked CSM (forward shock; Chevalier \& Fransson 1994). The lack of 
time-dependent X-ray spectroscopy precludes a distinction but, as we shall see, 
the X-ray lightcurve is consistent with the latter model with a constant 
pre-SN mass loss rate ($\rho_{\rm csm} \propto r^{-2}$). Alternatively, the CSM 
might result from a non-conservative mass-transfer to the companion star, leading to the 
formation of a flattened or disk-like H and He-rich shell (Nomoto et al. 1994). 
In this case a CSM profile flatter than $\rho_{\rm csm} \propto r^{-2}$ is expected 
due to the non-spherically symmetric geometry of the mass-transfer.

In the stellar wind scenario the continuum equation requires a mass-loss rate of 
$\dot{M} = 4\pi r^2 \rho_{\rm w}(r) \times v_{\rm w}(r)$ 
through a sphere of radius $r$. After the SN shock plows through the CSM, its density 
is $\rho_{\rm csm} = 4\rho_{\rm w}$ (Fransson, Lundqvist \& Chevalier 1996). 
The X-ray luminosity of the shock-heated CSM is 
$L_{\rm x} = \Lambda(T) {\rm d}V n^2$, where d$V$ is the volume, 
$n = \rho_{\rm csm}/m$ is the number density of the shocked CSM and $m$ is the 
mean mass per particle ($2.1\times10^{-27}$~kg for a H+He plasma). 
We thus obtain 
$L_{\rm x} = 4/(\pi m^2) \Lambda(T) \times (\dot{M}/v_{\rm w})^2\times(v_{\rm s} t)^{-1}$.
The observed X-ray rate of decline ($t^{-1}$) is consistent with this description
if $\Lambda(T)$, $\dot{M}/v_{\rm w}$ and $v_{\rm s}$ are constant.
We can hence use the observed X-ray luminosity at time $t$ after the outburst to measure 
the ratio $\dot{M}/v_{\rm w}$ assuming a constant shell expansion velocity $v_{\rm s}$. 
Given our observed X-ray luminosities (Table~1) and assuming a shell expansion velocity of 
$v_{\rm s} = 16,500~{\rm km~s}^{-1}$ (Filippenko et al. 1995) we derive a mass-loss rate 
of $\dot{M}\sim1\times10^{-5} M_{\odot}~{\rm yr}^{-1}~(v_{\rm w}/10~{\rm km~s}^{-1})$ 
consistent with all three detections and the \R\ upper limit on day 1,368. 

It should be noted, however, that a single power-law model can only be
reconciled with the data obtained on days 82--2,639, but not with the 
early \R\ upper limit on day 52 (see Fig.~2). This is indicative that
the above model might be incomplete for early epochs. The addition of an 
exponential term describing the rise during the early phase of the emission
leads to a fit that is in agreement with all measurements in case the
\R\ detection on day 82 was a post-maximum measurement ($s\sim1$). 
By contrast, the model with a steeper rate of decline ($s\sim1.5$), 
which postulates that the early \R\ detection corresponds to a 
pre-maximum measurement, is in conflict with a $\rho_{\rm csm} \propto r^{-2}$
profile. 

Table~1 summarizes the CSM number density $n$ for the different radii 
$r = v_{\rm s} \times t$ corresponding to the dates of the observations. 
As was already claimed based on the early \R\ data (Immler, Pietsch \& 
Aschenbach 1998), the $3\sigma$ upper limit for the CSM density at 
$r = 2.0\times10^{17}~{\rm cm}$ (day 1,368) is lower than that at
$1.2\times10^{16}~{\rm cm}$ (day 82). Our \C\ results at $3.2$ and 
$3.8\times10^{17}~{\rm cm}$ (days 2,271 and 2,639, respectively) 
clearly support these early indications for a decreasing CSM density 
profile (see Table~1). The combined \R\ and \C\ data of SN 1994I give 
a best-fit profile of $\rho_{\rm csm} \propto r^{-1.9\pm0.1}$.

The only SN for which a CSM density profile has been constructed from X-ray 
measurements is SN~1993J (Type IIb with a progenitor of $\sim15 M_{\odot}$), 
based on long-term monitoring with \R\ 
between six days and five years after the outburst (Immler, Aschenbach \& Wang 2001). 
A comparison between the CSM 
profiles of the two SNe is presented in Fig.~3. Although the CSM density profile of 
SN~1993J is significantly flatter ($\rho_{\rm csm} \propto r^{-1.6}$) 
than that of SN~1994I, the two SNe have a similar CSM number density of 
$\sim10^{6.5}~{\rm cm}^{-3}$ at a radius of $r\sim5\times10^{15}~{\rm cm}$ from the 
site of the explosion (cf. Fig.~3). 

Given that both the X-ray rate of decline ($L_{\rm x} \propto t^{-1}$) and
the CSM profile ($\rho_{\rm csm} \propto r^{-1.9\pm0.1}$) of SN~1994I are 
not in conflict with what is expected for a constant stellar wind speed and 
a constant mass-loss rate of the progenitor ($L_{\rm x} \propto t^{-1}$, 
$\rho_{\rm csm} \propto r^{-2}$), it is likely that the stellar wind of the massive 
progenitor dominates the CSM.

\acknowledgments

We thank Roger Chevalier for helpful discussion and the referee (Bernd Aschenbach) 
for his comments that led to significant improvement of the manuscript.
This research was supported by NASA grants NAG~5-8999 to the University 
of Massachusetts and NAG~81755 to the University of Maryland. 
Y.T. is supported by the Japan Society for the Promotion of Science
Postdoctoral Fellowship for Young Scientists.


\begin{deluxetable}{ccccccc}
\tabletypesize{\footnotesize}
\tablecaption{X-ray Properties of SN~1994I \label{tab}}
\tablewidth{0pt}
\tablehead{
\colhead{\phantom{0}Day$~^{\rm (a)}$} &
\colhead{Instrument} &
\colhead{Count Rate} &
\colhead{$f_{\rm x}~^{\rm (b)}$} &
\colhead{$L_{\rm x}~^{\rm (c)}$} &
\colhead{$\dot{M}~^{\rm (d)}$} &
\colhead{log$~n^{\rm (e)}$} \\
& & 
\colhead{($10^{-4}~{\rm cts~s}^{-1}$)} &
\colhead{($10^{-15}$)} &
\colhead{($10^{37}$)} &
\colhead{($10^{-5}$)} &
\colhead{(${\rm cm}^{-3}$)}}
\startdata
\phantom{0}\phantom{0}\,52 &
\R\ HRI &
$<3.1$ &
$<12.9$ &
$<10.9$ &
$<0.5$ &
$<5.9$ \\
\phantom{0}\phantom{0}\,82 &
\R\ HRI &
$5.2\pm1.7$ &
$21.6\pm5.0$ &
$18.2\pm4.2$ &
$0.7\pm0.4$ &
\phantom{0}\phantom{0}$5.7$ \\
1,368 &
\R\ HRI &
$<8.4$ &
$<34.8$ &
$<29.4$ &
$<3.8$ &
$<4.0$ \\
$2,271$ &
\C\ ACIS-S &
$5.5\pm2.2$ &
$1.7\pm0.7$ &
$1.4\pm0.5$ &
$1.1\pm0.7$ &
\phantom{0}\phantom{0}$3.0$ \\
2,639 &
\C\ ACIS-S &
$4.6\pm1.6$ &
$1.4\pm0.5$ &
$1.2\pm0.4$ &
$1.1\pm0.7$ &
\phantom{0}\phantom{0}$2.9$
\enddata
\tablenotetext{(a)}{\phantom{0}day after the outburst of SN~1994I
(March 31, 1994; Chandler, Phillips \& Rupen 1994)}
\tablenotetext{(b)}{\phantom{0}0.3--2~keV band flux in units of 
$10^{-15}~{\rm ergs~cm}^{-2}~{\rm s}^{-1}$ for a $10^7$~K thermal plasma
spectrum and a Galactic foreground column density of 
$N_{\rm H} = 1.3 \times 10^{20}~{\rm cm}^{-2}$ (Dickey \& Lockman 1990)}
\tablenotetext{(c)}{\phantom{0}0.3--2~keV band luminosity in units of 
$10^{37}~{\rm ergs~s}^{-1}$ for an assumed distance of $d = 8.4~{\rm Mpc}$
(Feldmeier, Ciardullo \& Jacoby 1997)}
\tablenotetext{(d)}{\phantom{0}pre-SN mass-loss rate in units of
$10^{-5}~M_{\odot}~{\rm yr}^{-1}$}
\tablenotetext{(e)}{\phantom{0}post-shock CSM number density}
\end{deluxetable}
\vfill


\begin{figure}[t!]
\centerline{ {\hfil\hfil
\psfig{figure=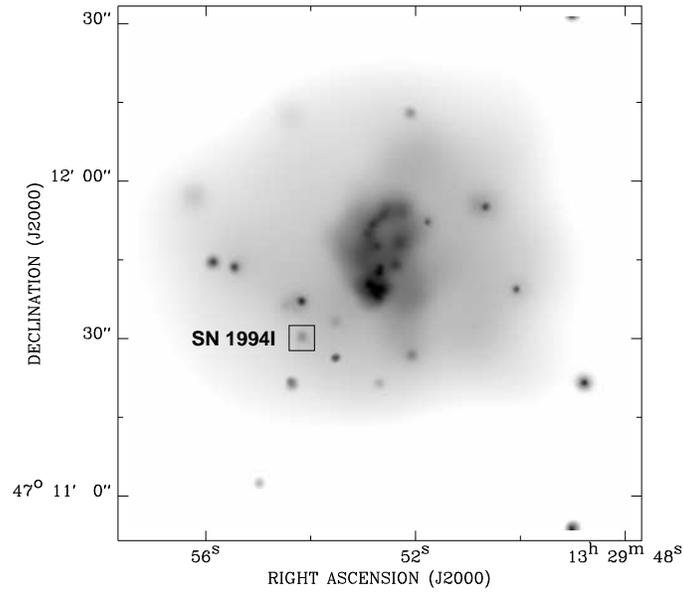,width=9cm,angle=270,clip=}
\hfil\hfil} }
\caption{\C\ ACIS-S soft (0.3--2~keV) band X-ray image of the central 
region of M51. The image was adaptively smoothed to achieve a signal-to-noise 
ratio in the range 3--5 and is plotted in logarithmic greyscale. 
The position of SN~1994I is marked by a box and the nucleus of M51 is at the center 
of the image.
\label{f1}}
\end{figure}

\begin{figure}[t!]
\centerline{ {\hfil\hfil
\psfig{figure=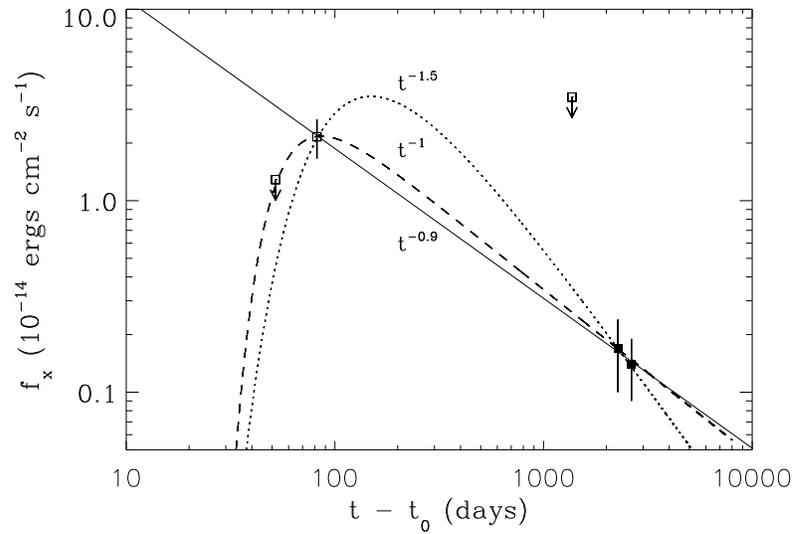,width=11cm,clip=}
\hfil\hfil}}
\caption{Soft (0.3--2~keV) band X-ray lightcurve of SN~1994I. 
The \C\ detections are marked by filled boxes, open boxes indicate the 
early \R\ HRI detection and $3\sigma$ upper limits. 
Error bars are $\pm1\sigma$ statistical errors.
Time is given in days after the outburst. 
The solid line represents a power-law fit to the
detections at days 82, 2,271 and 2,639. The dashed and dotted lines 
represent models including an $e^{-\tau}$ term (see Section~3).
\label{f2}}
\end{figure}

\begin{figure}[t!]
\centerline{ {\hfil\hfil
\psfig{figure=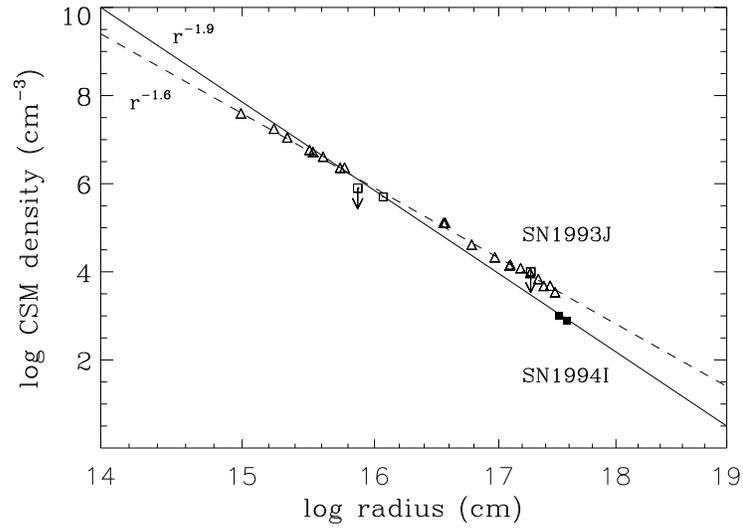,width=11cm,clip=}
\hfil\hfil} }
\caption{Circumstellar matter density profile as a function of SN shell 
expansion radius. The solid line gives the best-fit CSM density profile of 
$\rho_{\rm csm} \propto r^{-1.9}$ for the \R\ (open boxes) and \C\ (filled boxes) 
measurements of SN~1994I.
The CSM density profile with $\rho_{\rm csm} \propto r^{-1.6}$ 
based on \R\ observations of SN~1993J is drawn for comparison (dashed line and
open triangles; Immler, Aschenbach \& Wang 2001).
\label{f3}}
\end{figure}

\vfill

\end{document}